# Black Hole Constraints on Varying Fundamental Constants

Jane H. MacGibbon
*Dept of Physics and Chemistry, University of North Florida, Jacksonville Florida 32224, USA*

We apply the generalized second law of thermodynamics and derive upper limits on the variation in the fundamental constants. The maximum variation in the electronic charge permitted for black holes accreting and emitting in the present cosmic microwave background corresponds to a variation in the fine structure constant of $\Delta\alpha/\alpha \approx 2\times 10^{-23}$ per second. This value matches the variation measured by Webb et al. [*Phys.Rev.Lett.* **82**, 884-887 (1999); *Phys.Rev.Lett.* **87**, 091301 (2001)] using absorption lines in the spectra of distant quasars and suggests the variation mechanism may be a coupling between the electron and the cosmic photon background.

Observations [1,2] of absorption in the spectra from distant quasars raise the possibility that the fine-structure constant $\alpha$, which governs electromagnetic interactions, may be increasing as the Universe ages. The observations are consistent with a rate of change of roughly $\Delta\alpha/\alpha \approx 2\times 10^{-23}$ per second. The fine structure constant $\alpha = e^2/\hbar c$ where $e$ is the charge of the electron, $\hbar$ is Planck's constant and $c$ is the speed of light. Davies et al. [3] have proposed using black hole thermodynamics to limit the variation in the electronic charge $e$. In this Letter, we include the full description of the time variation of the entropy of the black hole system. We shall see that a small increase in $e$ of $\Delta e/e = \Delta\alpha/2\alpha \approx 10^{-23}$ per second does not violate the generalized entropy law for black holes in the present Universe. Thus black hole thermodynamical constraints do not rule out the possibility that an increase in $\alpha$ is due solely to an increase in electric charge $e$. Furthermore we will discover that $de/dt \approx 10^{-23} e$ per second matches the maximum variation in $e$ permitted for black holes in the present cosmic microwave background. Throughout this paper we assume that $c$, $\hbar$ and the gravitational constant $G$ are constant and investigate variation in $e$. Extension of this methodology to dependent or independent variation in the other fundamental constants is straightforward and will be presented elsewhere.

The generalized second law of thermodynamics, derived for black hole systems, states that the net entropy of the system cannot decrease with time [4]. Over a time interval $\Delta t$, the net generalized entropy of the system increases by

$$\Delta S = \Delta S_{BH} + \Delta S_{R+M} \geq 0, \quad (1)$$

where $\Delta S_{BH}$ and $\Delta S_{R+M}$ are the change in entropy of the black hole and of the ambient radiation and matter, respectively. The entropy of a black hole is [5,6]

$$S_{BH} = \frac{kc^3}{4\hbar G} A_{BH}, \quad (2)$$

where $k$ is the Boltzmann constant. For a charged, non-rotating (Reissner-Nordstrøm) black hole of mass $M$ and charge $Q$ in electrostatic (esu) units, the area of the black hole is

$$A_{BH} = \frac{4\pi G^2}{c^4}\left(M + \sqrt{M^2 - Q^2/G}\right)^2. \quad (3)$$

Hawking [5,6] has established that a black hole is continuously emitting quasi-thermal radiation with a temperature $T_{BH} = 2\hbar G\sqrt{M^2 - Q^2/G}/kcA_{BH}$. Thus $\Delta S_{BH}$, the full change in black hole entropy over time $\Delta t$, must include the contribution from the Hawking flux as well as any partial change induced by a variation in the electronic charge, i.e.

$$\Delta S_{BH} \approx \frac{dS_{BH}}{dt}\Delta t = \frac{kc^3}{4\hbar G}\left(\frac{\partial A_{BH}}{\partial t} + \frac{\partial A_{BH}}{\partial e}\frac{de}{dt}\right)\Delta t. \quad (4)$$

Only the second term was considered by Davies et al. [3]. In the general case,

$$\frac{\partial A_{BH}}{\partial t} = \frac{8\pi G^2}{c^4}\frac{\left(M + \sqrt{M^2 - Q^2/G}\right)}{\sqrt{M^2 - Q^2/G}}$$
$$\times\left\{\left(M + \sqrt{M^2 - Q^2/G}\right)\frac{dM_H}{dt} - \frac{Q}{G}\frac{\partial Q_H}{\partial t}\right\} \quad (5a)$$

$$\frac{\partial A_{BH}}{\partial e} = \frac{8\pi G^2}{c^4}\frac{\left(M + \sqrt{M^2 - Q^2/G}\right)}{\sqrt{M^2 - Q^2/G}}\left\{-\frac{Q}{G}\frac{\partial Q}{\partial e}\right\}, \quad (5b)$$

where $\partial Q/\partial e = Q/e$ (if $\partial Q/Q = \partial e/e$) and the subscript $H$ denotes Hawking radiation (or where appropriate thermal accretion). Both $M$ and $Q$ change as the black hole radiates. To proceed further we must consider the two cases when the black hole temperature is greater than and less than the temperature of its surroundings.

Case (I).-If the black hole temperature is greater than the temperature of its surroundings, ie $T_{BH} > T_{R+M}$, there will be a net radiation loss from the black hole into its environment.

Case (IA).- Consider first the case when $Q$ is not affected by the Hawking radiation, i.e. the black hole temperature is below about 100 keV, the threshold to emit the lightest charged particle, the electron. This corresponds to $M \gtrsim 10^{17}$ g. Then $\partial Q_H / \partial t = 0$ and so

$$\frac{dS_{BH}}{dt} = \frac{2\pi kG}{\hbar c}\left[1+\sqrt{1-Q^2/GM^2}^{-1}\right]$$
$$\times \left\{\left(M+\sqrt{M^2-Q^2/G}\right)\frac{dM_H}{dt} - \frac{Q}{G}\frac{\partial Q}{\partial e}\frac{de}{dt}\right\}. \quad (6)$$

The mass loss due to Hawking radiation is [7]

$$\frac{dM_H}{dt} \approx -\frac{\hbar c^4}{G^2 M^2}\beta \quad (7)$$

with $\beta \approx 3 \times 10^{-4}$ for a hole emitting the photon, 3 light neutrino species and the graviton. Strictly the mass loss rate (7) applies only if $Q << Q_{MAX} = G^{1/2}M$, the maximal possible charge on a black hole. This will suffice for our purpose because, as described below, high $Q$ is discharged quickly provided that $M \lesssim \hbar c e / G^{3/2} m_e^2 \approx 10^5 M_\odot$ where $m_e$ is the electron mass [8-11]. Page [12] has numerically calculated that for a black hole emitting the photon, 3 light neutrino species and the graviton the increase in $S_{R+M}$ due to Hawking emission is 1.62 times the corresponding decrease in $S_{BH}$ due to Hawking emission. Thus $\Delta S \geq 0$ provided the second term within the { } brackets in Eq (6) is not of order the first term. The second term is only of order the first term when the black hole has a charge of

$$Q_{1\approx 2} \approx \left\{\frac{\hbar c^4 \beta}{GM\left(e^{-1}de/dt\right)}\right\}^{1/2}\left(2-\frac{\hbar c^4 \beta}{G^2 M^3 \left(e^{-1}de/dt\right)}\right)^{1/2}.$$
(8)

If $de/dt \approx 10^{-23} e$ per second, then $Q_{1\approx 2} \lesssim Q_{MAX}$ for $M \gtrsim 1.8 \times 10^{16}$ g and so $Q_{1\approx 2}$ could be achievable. However, Gibbons [9] and Zaumen [10] have shown that if the charge is greater than $Q_{PP} \approx G^2 m_e^2 M^2 / \hbar c e$, the black hole will quickly discharge by superradiant [13,14] Schwinger-type $e^+e^-$ pair-production in the electrostatic field surrounding the hole. From Eq (5a) this Schwinger-type discharge will increase $\partial A_{BH}/\partial t$ and hence $S_{BH}$, as well as $S_{R+M}$. The superradiant threshold $Q_{PP}$ is less than $Q_{1\approx 2}$ for all holes lighter than

$$M_{PP} \approx \left(\frac{2\hbar^3 c^6 e^2 \beta}{G^5 m_e^4 \left(e^{-1}de/dt\right)}\right)^{1/5} \quad (9)$$

i.e. $M_{PP} \approx 2.6 \times 10^{25}$ g for $de/dt \approx 10^{-23} e$ per second. Gibbons [9] has derived the discharge rate for $T_{BH} << 100\,\text{keV}$ black holes with $Q > Q_{PP}$ to be $dQ_{PP}/dt \approx (e^4 Q^3 / \hbar^3 c^2 r_+) \exp\left(-\pi c^3 m_e^2 r_+^2 / \hbar Q e\right)$, where $r_+ = G\left(M+\sqrt{M^2-Q^2/G}\right)/c^2$. Thus the pair-production discharge term $(Q/G)dQ_{PP}/dt$ from a $Q_{1\approx 2}$ black hole is greater than the entropy decrease term $(Q/G)(\partial Q/\partial e)de/dt$ in Eq (6) for all $M \lesssim M_{PP \approx E}$ where $M_{PP \approx E}$ satisfies

$$\frac{c^4 e^4 \beta}{\hbar^2 G^2 M^2 \left(e^{-1}de/dt\right)^2}$$
$$\approx \exp\left(\frac{4\pi G^{5/2} m_e^2 M^{5/2} \left(e^{-1}de/dt\right)^{1/2}}{\sqrt{2}\hbar^{3/2} c^3 e \beta^{1/2}}\right) \quad (10)$$

i.e. $M_{PP \approx E} \approx 7.0 \times 10^{25}$ g for $de/dt \approx 10^{-23} e$ per second. Because the pair-production term grows much faster with $Q$ than the $de/dt$ term, the pair-production term then dominates when $Q_{1\approx 2} \lesssim Q$ for $M \lesssim M_{PP \approx E}$. To the accuracy of our analysis and the original references (which together may be roughly ± a factor of 2) and the measurements of $\Delta\alpha/\alpha$, our estimate of $M_{PP,PP=E} \approx 2.6 - 7.0 \times 10^{25}$ g for $de/dt \approx 10^{-23} e$ per second coincides with $M_{CMB} \approx 4.5 \times 10^{25}$ g, the mass of a black hole whose temperature is equal to the ambient temperature of the Universe $T = 2.73\,\text{K}$.

Thus in case (IA), the net increase in $S_{BH} + S_{R+M}$ due to Hawking emission and pair production is greater than the decrease in $S_{BH}$ induced by an electronic charge change of $de/dt \approx 10^{-23} e$ per second for all neutral and charged black holes whose temperature is greater than the 2.73K cosmic microwave background.

Case (IB).- Consider the case when the black hole is emitting charged particles via Hawking emission, ie $\partial Q_H / \partial t \neq 0$ and $M \lesssim 10^{17}$ g. A charged black hole emits its charge at a rate which depends on $Q$ and preferentially emits particles of the same sign as its own charge. Then $\partial Q_H / \partial t = -(e|Q|/Q)dN_H/dt$, where $edN_H/dt$ is the net emission rate of charge out of the black hole. If $E_{av} \approx 5T_{BH}$ is the average energy of a particle emitted by the black hole [7] and $dN_{TOT}/dt$ is the total emission rate of all particles from the black hole, then $dM_H/dt = -(E_{av}/c^2)dN_{TOT}/dt \leq 0$ and $|Q\partial Q_H/\partial t| \leq (c^2e/E_{av})|QdM_H/dt|$. Thus $dS_{BH(IB)}/dt$ lies in the range

$$\frac{2\pi kG}{\hbar c}\left[1+\sqrt{1-Q^2/GM^2}\right]^{-1}$$
$$\times\left\{\left(M+\sqrt{M^2-Q^2/G}-\frac{8\pi M|Q|e}{5\hbar c}\right)\frac{dM_H}{dt}-\frac{Q}{G}\frac{\partial Q}{\partial e}\frac{de}{dt}\right\}$$
$$\geq \frac{dS_{BH(IB)}}{dt} \geq \frac{2\pi kG}{\hbar c}\left[1+\sqrt{1-Q^2/GM^2}\right]^{-1}$$
$$\times\left\{\left(M+\sqrt{M^2-Q^2/G}\right)\frac{dM_H}{dt}-\frac{Q}{G}\frac{\partial Q}{\partial e}\frac{de}{dt}\right\}. \quad (11)$$

For $Q \ll Q_{MAX}$ it is straightforward to show that the net entropy increase due to the Hawking emission, given by Eq (7) with $\beta \approx 4 \times 10^{-4}$ for $M \lesssim 10^{17}$ g [15], dominates the entropy decrease due to $de/dt \approx 10^{-23}e$ per second. For higher $Q$, the relevant Hawking emission rate per degree of particle freedom of spin $s$ particles with energy in the range $(E, E+dE)$ is

$$d\dot{N} = \frac{\Gamma_s dE}{2\pi\hbar}\left\{\exp\left(\frac{E-c^2eQ/GM}{kT_{BH}}\right)-(-1)^{2s}\right\}^{-1} \quad (12)$$

where $\Gamma_s$ is the spin- and charge-dependent absorption probability. As $Q$ increases, the emission rate is modified by the electrostatic chemical potential term in Eq (12). Carter [11] has estimated that $dQ_e/dt \approx -c^2e^2Q/\hbar GM$ for $M \lesssim 10^{17}$ g and $Q \leq \hbar c/e$ (the thermal regime), and $dQ_e/dt \approx -e^4Q^3/\hbar^3 GM$ for $Q \geq \hbar c/e$ (the superradiant regime) which matches the superradiant discharge rate for larger $M$. In both cases the entropy increase due to the $dQ_e/dt$ term dominates the entropy decrease due to $de/dt \approx 10^{-23}e$ per second for all $M \lesssim 10^{17}$ g. (We also note that the discharge timescale $\tau_Q \approx Q/\dot{Q}_e$ for a $M \lesssim 10^{17}$ g hole is much smaller than its lifetime $\tau_{BH} \approx G^2M^3/\hbar c^4$ and is even comparable with or less than $\hbar c/e^2 = 137$ times the characteristic timescale it takes to form $\tau_F \approx r_+/c$ [11], so $M \lesssim 10^{17}$ g black holes should be essentially neutral today [8,9], up to random fluctuations of order the Planck charge $(\hbar c)^{1/2}$.) Thus for all neutral and charged black holes in case (IB), there is a net increase in $S_{BH} + S_{R+M}$ if $de/dt \approx 10^{-23}e$ per second.

Case (II).-If the black hole temperature is less than or equal to the temperature of its surroundings, i.e. $T_{BH} \leq T_{R+M}$, the net entropy also increases when $e$ increases at the rate indicated by the Webb et al. observations [1]. This can be shown by explicitly deriving the heat flow into the hole or by general thermodynamical principles as follows.

An increase in $e$ will decrease $A_{BH}$ and $T_{BH}$. Once $T_{BH}$ drops below the ambient temperature, the black hole will accrete from its surroundings faster than it Hawking radiates. This accretion increases the black hole mass $M$, further lowering $T_{BH}$, and leads in turn to more accretion. (As Hawking has pointed out [16], a black hole cannot be in stable thermal equilibrium if an unbounded amount of energy is available in its surroundings. This also means that the Davies et al.[3] suggestion that a black hole can be kept in isoentropic equilibrium with a same temperature heat bath is not achievable.) The general thermodynamical definitions of the temperature of the environment and the black hole temperature are, respectively, [16]

$$T_{R+M}^{-1} \equiv \frac{\partial S_{R+M}}{\partial E}, \quad T_{BH}^{-1} \equiv c^{-2}\left(\frac{\partial S_{BH}}{\partial M}\right)_{Q\text{ fixed}}, \quad (13)$$

where $E$ is the energy of the environment. During accretion, the black hole mass increases by an amount equal to the decrease in $E$. Hence for $T_{BH} \leq T_{R+M}$, the temperature definitions imply that the increase in black hole entropy due to accretion must be greater than the decrease in $S_{R+M}$ due to accretion. Also for $T_{BH} \leq T_{R+M}$, the increase in $S_{BH}$ due to accretion must be greater than the decrease in $S_{BH}$ due to Hawking radiation. In analogy with a classical blackbody, a cold large black hole in a warm thermal bath will absorb energy at a rate $dE/dt = \pi^2 k^4 \sigma_S T_{R+M}^4/60\hbar^3 c^2$ (and emit radiation $dE/dt = \pi^2 k^4 \sigma_S T_{BH}^4/60\hbar^3 c^2$) per polarization or helicity eigenstate where $\sigma_S = 27\pi G^2 M^2/c^4$ is the geometrical optics cross-section [7]. (Since the entropy of the background is maximized for a thermal bath, a thermal bath will give the strictest accretion constraint on $\Delta S$.) For accretion, Eq (7) is thus replaced by

$$\frac{dM}{dt} = +\frac{\beta_{R+M}\hbar c^4}{G^2 M_{R+M}^2}\left(\frac{M}{M_{R+M}}\right)^2, \quad (14)$$

where $\beta_{R+M} \approx 10^{-4}$ and $M_{R+M}$ is the mass of a black hole whose temperature equals the ambient temperature. The $de/dt$ term in Eq (6) is now of order the first (absorption) term only when

$$Q'_{1\approx 2} = \left\{\frac{\hbar c^4 \beta_{R+M}}{GM_{R+M}\left(e^{-1}de/dt\right)}\right\}^{1/2}\left(\frac{M}{M_{R+M}}\right)^{3/2}$$
$$\times\left(2-\frac{\hbar c^4 \beta_{R+M}}{G^2 M_{R+M} M^2\left(e^{-1}de/dt\right)}\right)^{1/2}. \quad (15)$$

Provided $M \lesssim 5 \times 10^{20}\left(M_{R+M}/M_{CMB}\right)^4 M_\odot$, then

$Q_{MAX} \lesssim Q'_{1\approx 2}$ for $de/dt \approx 10^{-23} e$ per second. Gibbons [9] has argued that large black holes should not acquire significant charge or approach $Q_{MAX}$. Once $Q/G^{1/2}M > G^{1/2}m_e/e \approx 5\times 10^{-22}$, a black hole can only gravitationally accrete a particle of like charge if the particle is projected at the black hole with an initial velocity [9] and a large black hole would be more likely to lose charge by accreting a particle of opposite charge. More rigorously, the generalized third law of thermodynamics states that $T_{BH} = 0$, and hence $Q_{MAX}$, is not achievable by a finite sequence of steps [12,17]. Thus the black hole charge will remain below $Q_{MAX}$ and the accretion from the radiation background, which depends on $T_{R+M}^4$, will always dominate the $de/dt$ term. Addressing $M \gtrsim 5\times 10^{20} (M_{R+M}/M_{CMB})^4 M_\odot$, such supermassive black holes can not exist in the present Universe. Such a black hole would have a Schwarzschild radius of 25 – 50 Mpc, which is at least 1% of the current cosmic horizon, and would be at least 10 orders of magnitude more massive than an active galactic nuclei core. Additionally, a black hole can only form when the size of the Universe is greater than the Schwarzschild radius of the hole and any black hole should produce noticeable distortion in its surrounding space-time out to at least about 10 times its Schwarzschild radius. No distortion on the scale of 1% – 10% of the cosmic horizon is observed today. Stated another way, the existence of such supermassive black holes is ruled out by the present age and structure of the Universe. However even if such a supermassive black hole did exist it would presumably also discharge quickly by charge accretion from its environment. Thus $Q'_{1=2}$ is not attainable by any black hole in the present Universe and even in the ultra-massive limit $\Delta S \geq 0$ is not violated.

In the special case when $T_{R+M} \geq T_{BH} \geq T_{CMB}$, $\Delta S$ due to absorption again must be greater than $\Delta S$ due to emission and, as we have shown in case (I), $\Delta S$ due to emission is greater than or equal to the decrease in entropy due to $de/dt \approx 10^{-23} e$ per second for $T_{BH} \geq T_{CMB}$. Hence for all $T_{BH} \leq T_{R+M}$, an increase in $e$ of the size indicated by the Webb et al. data [1] must produce a net increase in the generalized entropy, i.e. $\Delta S_{BH} + \Delta S_{R+M} \geq 0$ for all black holes with a net absorption.

Combining cases (IA), (IB) and (II), we conclude that if the electronic charge $e$ increases at a rate consistent with the Webb et al. observations, the generalized second law of thermodynamics is not violated by black holes in the present Universe. In fact our analysis shows that a change in the fine structure constant of $\Delta\alpha/\alpha \approx 2\times 10^{-23}$ per second corresponds to the maximum increase in $e$ allowed by the generalized second law of thermodynamics for black holes in the present Universe. The second law could be violated by emitting black holes if the Universe were only somewhat colder than today.

Extending our analysis to rotating black holes is straightforward and does not modify our conclusions. For a rotating, charged (Kerr-Newman) black hole, the $M + \sqrt{M^2 - Q^2/G}$ factor is replaced by $M + \sqrt{M^2 - Q^2/G - c^2 J^2/G^2 M^2}$. The maximal rotation is $J_{MAX} = GM^2 \sqrt{1 - Q^2/GM^2}/c$ and $J_{MAX} \approx GM^2/c$ unless $Q$ is very close to $Q_{MAX} = G^{1/2}M$ in which case the above treatment of $Q_{MAX}$ should be followed. Page [12] has shown that from $J = 0$ to $J = GM^2/c$, the power of a black hole emitting 4 spin-1/2 (3 neutrino species and electrons), 1 spin-1 (photon) and 1 spin-2 (graviton) species increases by a factor of 300 and the black hole loses spin faster than it loses energy. In the case of a $J \approx GM^2/c$ or $J = GM^2/c$ black hole, the superradiant mechanism for bosonic and fermionic modes dominates [12-14]. By the superradiant mechanism, which we discussed above for charged fermionic modes, a particle-antiparticle pair is created in the ergosphere with one particle with positive energy escaping to infinity and the other particle with locally positive energy being absorbed by the black hole. This mechanism has the consequence of increasing both the entropy of the environment and the entropy of the black hole, and spinning down the extremal black hole. In the case of a non-extremal black hole, the generalized third law of thermodynamics can also be applied to show that an existing black hole can not be spun up to $J_{MAX}$. Therefore the strictest constraints we obtain by including the $de/dt$ term in the generalized second law of thermodynamics come from black holes with $J = 0$ and charge $Q > 0$. Our conclusions are also not affected by the changes due to $de/dt$ in the Hawking and pair production discharge rates which are second order effects.

It should be noted that our derivation is essentially standard model physics and does not invoke quantum gravity. The black hole entropy and temperature, as defined, are required for *classical* general relativity to be consistent with *classical* thermodynamics [16,18,19]. Additionally the superradiant mechanism was first described by Zel'dovich [13,14] for classical black holes prior to the discovery of Hawking radiation. Schwinger pair production [20] is a non-perturbative process in standard QED.

If the Webb et al. measurements [1] are correct, our analysis suggests at least two possibilities. We postulate that nature is such that $e$ varies at the maximal rate allowed by the generalized second law of thermodynamics. If this is so then, as seen in Eqs (9) and (10), the rate of increase in $\alpha$ should weaken with time as the Universe cools and $M_{CMB}$ increases. This postulate could be expanded if the increase in $\alpha$ is due not solely to $e$ varying, but to $e$, $\hbar$, $c$ and/or $G$ varying dependently as proposed in some standard model extensions, to say that the combined variation occurs at the maximal rate allowed by the generalized second law of thermodynamics. The maximal variation postulate should be explored theoretically and experimentally.

Additionally, the form of our results strongly suggests that, if the Webb et al. measurements [1] are correct, the increase in $\alpha$ may be due to a higher order coupling between the electron charge and the cosmic photon background whose effect is to partially screen the bare electron charge. As the Universe cools, the coupling weakens, increasing $e$. Although we have derived our result by applying the generalized second law of thermodynamics to black holes, in doing so we may have mathematically mimicked the relevant cosmological calculation or rather, derived it from a Principle instead of the explicit details of the mechanism: since the photon background is cosmological in origin its temperature implicitly depends on $G$. It should be investigated whether such a coupling arises as a higher order effect in standard QED or in standard model extensions. Because our strongest black hole constraint comes from the Schwinger pair production regime, an obvious candidate mechanism to consider is the scattering of the vacuum polarization $e^+e^-$ around a bare electron off the cosmic photon background.

It is a pleasure to thank the University of Cambridge for hospitality.